\documentclass[twocolumn,aps,prb,superscriptaddress,showkeys]{revtex4}

\usepackage{amssymb}
\usepackage{amsmath}
\usepackage{amsfonts}
\usepackage{graphicx}

\begin{document}

\title{Spin-dephasing anisotropy for electrons in a diffusive quasi-1D GaAs wire}
\author{J.~Liu}
\author{T.~Last}
\altaffiliation[Electronic mail:]{ t.last@rug.nl}
\altaffiliation[Phone:]{ +31(0)503638974}
\author{E.~J.~Koop}
\author{S.~Denega}
\author{B.~J.~van~Wees}
\author{C.~H.~van~der~Wal}
\affiliation{Physics of Nanodevices Group, Zernike Institute for Advanced
Materials, University of Groningen, Nijenborgh 4, NL-9747AG
Groningen, The Netherlands}

\date{\today}

\begin{abstract}
We present a numerical study of dephasing of electron spin ensembles
in a diffusive quasi-one-dimensional GaAs wire due to the
D'yakonov-Perel' spin-dephasing mechanism. For widths of the wire
below the spin precession length and for equal strength of Rashba
and linear Dresselhaus spin-orbit fields a strong suppression of
spin-dephasing is found. This suppression of spin-dephasing shows a
strong dependence on the wire orientation with respect to the
crystal lattice. The relevance for realistic cases is evaluated by
studying how this effect degrades for deviating strength of Rashba
and linear Dresselhaus fields, and with the inclusion of the cubic
Dresselhaus term.
\end{abstract}





\maketitle

\section{\label{sec:Introduction}INTRODUCTION}
\begin{figure}[tbh]
\begin{center}
\includegraphics[width=8cm]{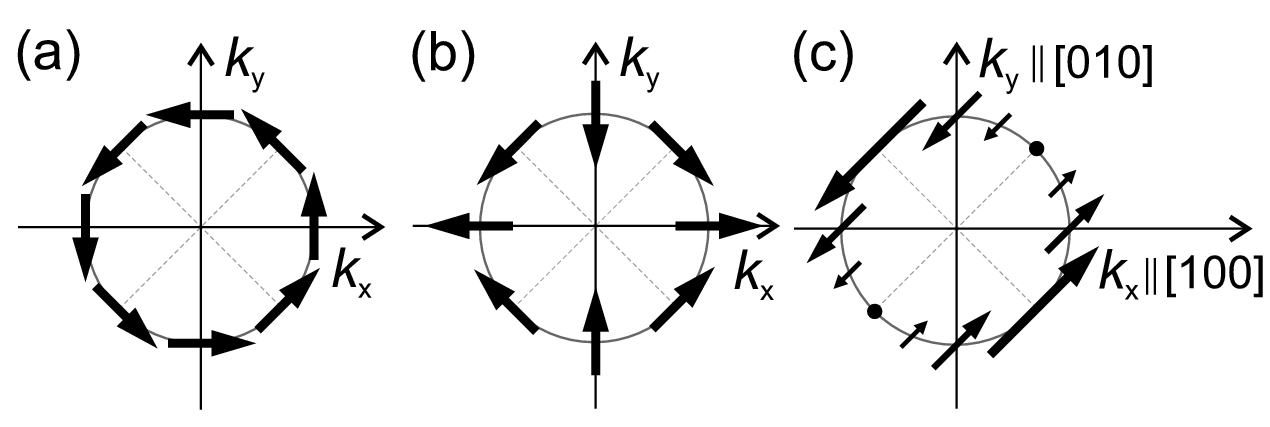}
\end{center}
\caption{A schematic representation of the direction and magnitude
of the effective magnetic field for selected points in a
two-dimensional $k$-space, sketched for (a) the Rashba field, (b)
the linear Dresselhaus field and for (c) the symmetric case of the
sum of equal Rashba and linear Dresselhaus field. Both the magnitude
and direction are depicted as arrows on the Fermi circle with radius
$k_F$ in the ($k_x$, $k_y$)-plane.} \label{1}
\end{figure}

The ability to maximize the spin-dephasing time $T_{2}^*$
of an electron spin ensemble is one of the key issues for
developing semiconductor-based spintronic devices
\cite{review, Kikkawa1998}. However, in all III-V
semiconductor materials spin ensembles rapidly dephase due
to the D'yakonov-Perel' (DP) spin-dephasing mechanism
\cite{D'yakonov1986, Miller2003}. For the case of electron
ensembles in a heterojunction two-dimensional electron gas
(2DEG), two distinct contributions to DP spin-dephasing
have to be considered: the inversion asymmetry of the
confining potential (structural inversion asymmetry) and
the bulk inversion asymmetry of the crystal lattice. The
former results in an effective Rashba field and the latter
in an effective Dresselhaus field, which includes linear
and cubic contributions \cite{D'yakonov1986,Bychkov1984,
Lommer1985,Miller2003}:
\begin{eqnarray}
\vec{B}_{R} \ &=& \ C_{R}~(\hat{x}k_{y}-\hat{y}k_{x}),
\\
\vec{B}_{D1} \ &=& \ C_{D1}(-\hat{x}k_{x}+\hat{y}k_{y}),
\\
\vec{B}_{D3} \ &=& \ C_{D3}(\hat{x}k_xk_{y}^2-\hat{y}k_{x}^2k_y),
\end{eqnarray}
where $\hat{x}$, $\hat{y}$ are the unit vectors along the [100] and
[010] crystal directions, $k_{x}$, $k_{y}$ are the components of the
in-plane wave vector, and $C_{R,D1,D3}$ are the spin-orbit coupling
parameters. The total effective spin-orbit field $\vec{B}_{eff}$ is
the vector sum of all three contributions. For 2D and quasi-1D
electron systems, the direction and magnitude of these effective
spin-orbit fields can be illustrated as arrows on the Fermi circle.
Figure~\ref{1} presents this for selected points in the 2D momentum
space, for the Rashba (a) and linear Dresselhaus (b) field alone,
and their sum (c) for the case of equal strength of Rashba and
linear Dresselhaus field. In contrast to the individual cases
(Figure~\ref{1} (a), (b)), the magnitude of the vector sum shows a
strong anisotropy in momentum space (Figure~\ref{1} (c)).
This already suggests that spin dephasing in very narrow wires in
which electron motion is restricted to the [110] direction can be
strongly reduced as compared free 2DEG or such wires oriented along
other crystal directions. However, it is harder to analyze whether
such a dephasing anisotropy also occurs for wider quasi-1D wires,
where the motion in the 2DEG plane is still completely random and
diffusive, but where the width of the wire is less than length
scales as the spin precession length or the mean free path. For the
latter case the transport regime could be named quasi-ballistic, but
we consider the case of a large ensemble where transport along the
wire is still diffusive, and where the width of the wire in the 2DEG
plane is much wider than the regime of quantum confinement. Initial
studies of such spin-dephasing anisotropy include a recent
experiment \cite{Folk2008} on wires, and theoretical work on
drifting ensembles in free 2DEG \cite{Loss2007}. However, until now
most emphasis was on work related to the spin field-effect
transistor \cite{Datta1990}, using InAs-based systems or highly
asymmetrical heterojunctions, where structural inversion asymmetry
dominates the spin-orbit interaction \cite{Kiselev2000, Bruno2002,
Mireles2001, Pramanik2003, Holleitner2006}.
\\
We report here how the D'yakonov-Perel' spin dephasing mechanism can
be strongly suppressed in diffusive quasi-1D electron systems based
on GaAs heterojunction material, for which Rashba and linear
Dresselhaus spin-orbit contributions can be indeed of comparable
magnitude \cite{Miller2003,Folk2008}. The dephasing is studied for
spin ensembles initially aligned perpendicular to the plane of the
wire ([001] direction). This situation reflects the method of
preparing and interrogating a spin population via optical pump-probe
techniques \cite{Kikkawa1998}. Our numerical calculation is first
performed for conditions with equal Rashba and linear Dresselhaus
contributions and the cubic Dresselhaus term set to zero. For widths
of the diffusive quasi-1D wires smaller than the spin precession
length the DP spin-dephasing mechanism can be strongly suppressed
and the spin-dephasing time $T_2^*$ is considerably enhanced if the
wire is aligned along the direction of zero effective spin-orbit
field. Moreover, we want to point out that the value of our
numerical tool lies in the opportunity to study such phenomena also
for more realistic conditions. Thus, we can study how breaking the
equality of the Rashba and linear Dresselhaus spin-orbit fields, or
adding the cubic Dresselhaus term leads to a degradation of the
spin-dephasing anisotropy.

\section{\label{Method}Method}
We apply a Monte Carlo method \cite{Koop2008} to study the
temporal evolution of the normalized spin orientation
(average spin expectation value) in an elongated quasi-1D
wire. Our numerical tool is based on a semiclassical
approach. We use a classical description for the electron
motion, and a quantum mechanical description of the
dynamics of the electron spin. The wire is treated as a
rectangular box of aspects 1 $\mu$m and 200 $\mu$m. The
electron density and mobility are set to 4$\cdot10^{15}$
m$^{-2}$ and 100 m$^{2}$/Vs, which are typical values for a
GaAs/AlGaAs heterojunction material. All electrons are
assumed to have the same Fermi velocity $\upsilon_{F}$ of
2.7$\cdot10^{5}$ m/s. This is a valid approximation for
$k_BT$, $\Delta E_{Z,SO}$ $\ll$ $E_F$ (with respect to the
bottom of the conduction band), where $\Delta E_{Z,SO}$ is
the Zeeman splitting due to the spin-orbit fields alone.
Electron-electron interaction and inelastic scattering
mechanisms are neglected.
\\
The electron is regarded as a point particle which moves on
a classical trajectory between scatter events on impurities
(randomly determined at a rate to obtain an average scatter
time of 38 ps) yielding diffusive behavior in the ensemble
(electron mean free path $L_{p}$ = $\upsilon_{F}\tau_{p}$ =
10~$\mu$m), and specular scattering on the edges of the
wire. For each electron moving on such a ballistic
trajectory we calculate the spin evolution in the effective
spin-orbit fields quantum mechanically, and we then take
the ensemble average on a set of electrons with random
initial position and momentum direction.
\\
Within a straight ballistic segment of an electron
trajectory the spin rotates around $\vec{B}_{eff}$ over a
precession angle given by
\begin{eqnarray}
\phi_{prec} \ = \ \frac{g \mu_B |\vec{B}_{eff}|}{\hbar} \ t,
\end{eqnarray}
where $\hbar$ is the reduced Planck's constant and $t$ the
time of traveling through the segment. The spin rotation
operator $\widehat{U}$ for rotation over the precession
angle $\phi_{prec}$ about the direction $\vec{u}$ (unit
vector) of the effective magnetic field $\vec{B}_{eff}$ is
obtained by (see e.g. \cite{Claude})
\begin{widetext}
\begin{eqnarray}
\widehat{U} \ = \ \text{exp}(-i \ \frac{\phi_{prec}}{2} \
\vec{\sigma} \ \vec{u}) \ = \ \left(
\begin{array}{*{2}{c}}
\text{cos} \frac{\phi_{prec}}{2}-iu_z\text{sin} \frac{\phi_{prec}}{2} & -(iu_x+u_y)\text{sin}\frac{\phi_{prec}}{2}  \\
-(iu_x-u_y)\text{sin}\frac{\phi_{prec}}{2} & \text{cos}
\frac{\phi_{prec}}{2}+iu_z\text{sin} \frac{\phi_{prec}}{2} \\
\end{array}
\right),
\end{eqnarray}
\end{widetext}
where $\vec{\sigma}$ represents the vector of Pauli spin
matrices (x, y, z components of the spin). $\widehat{U}$
acting on a spin state $| \Psi_{initial} \rangle$ at the
beginning of a ballistic trajectory yields the spin state $
| \Psi_{final} \rangle \ = \widehat{U} \ | \Psi_{initial}
\rangle$ at the end of the trajectory. Thus, we can follow
the spin state of each electron (labeled $i$), and we use
this to define a semiclassical spin vector to present its
orientation, $ \vec{S}_i = ( \langle S_x \rangle_i, \langle
S_y \rangle_i, \langle S_z \rangle_i ) $ from its spin
expectation values in $x$, $y$, $z$ directions.

For an electron experiencing multiple scattering events,
the orientation of the effective magnetic field is changed
at each scatter event. For each trajectory a rotation is
applied. Once a scattering event takes place, the wave
vector state is updated, $\vec{B}_{eff}$ is recalculated
based on the new wave vector, resulting in a new rotation
operator, and the evolution of the spin state will carry
on. For spin ensembles, the randomization will bring a
reduction of the normalized spin orientation (average spin
expectation value) $\langle S_r \rangle \ = \ \left| (
\sum_{i=1}^N \vec{S}_i )/ N \right|$ for the ensemble.
$\langle S_r \rangle$ is obtained by averaging over an
ensemble of $N$ = 1000 spins, independent of their
positions within the system, and calculated as a function
of time. We choose to study the spin coherence in the
ensemble here as $\langle S_r \rangle$. The advantage is
that $\langle S_r \rangle$ gives the magnitude of the
residual spin orientation in the direction that is maximum
(automatically evaluating the envelope in case the ensemble
average shows precession). However, in the present study
without externally applied fields, there was no development
of average spin orientation in the $x$ and $y$ directions,
and the decay of $\langle S_r \rangle$ always equaled the
decay of ensemble average $\langle S_z \rangle \ = \ \left|
( \sum_{i=1}^N \langle S_z \rangle_i )/ N \right|$. The
spin-dephasing time $T_2^*$ of the ensemble is defined as
the decay time over which $\langle S_r \rangle$ reduces to
1/e of its initial value. Note, however, that decay traces
of $\langle S_r \rangle$ were not always mono exponential
in our simulations.

\section{Results and Discussion}
The temporal evolution of the normalized spin orientation $\langle
S_r \rangle$ is studied for out-plane initial spin states oriented
along the [001] direction. The effective spin-orbit field resulting
from equal magnitudes of the Rashba and linear Dresselhaus fields is
always parallel to the [110] axis ($C_R$ = $C_{D1}$ =
$-1.57\cdot$10$^{-8}$ Tm, \cite{Miller2003}). These values of the
spin-orbit parameters give rise to an average spin precession length
of about 3 $\mu$m. The width of our wire of 1 $\mu$m is chosen to be
smaller than this length scale.
\begin{figure}[h]
\begin{center}
\includegraphics[width=8cm]{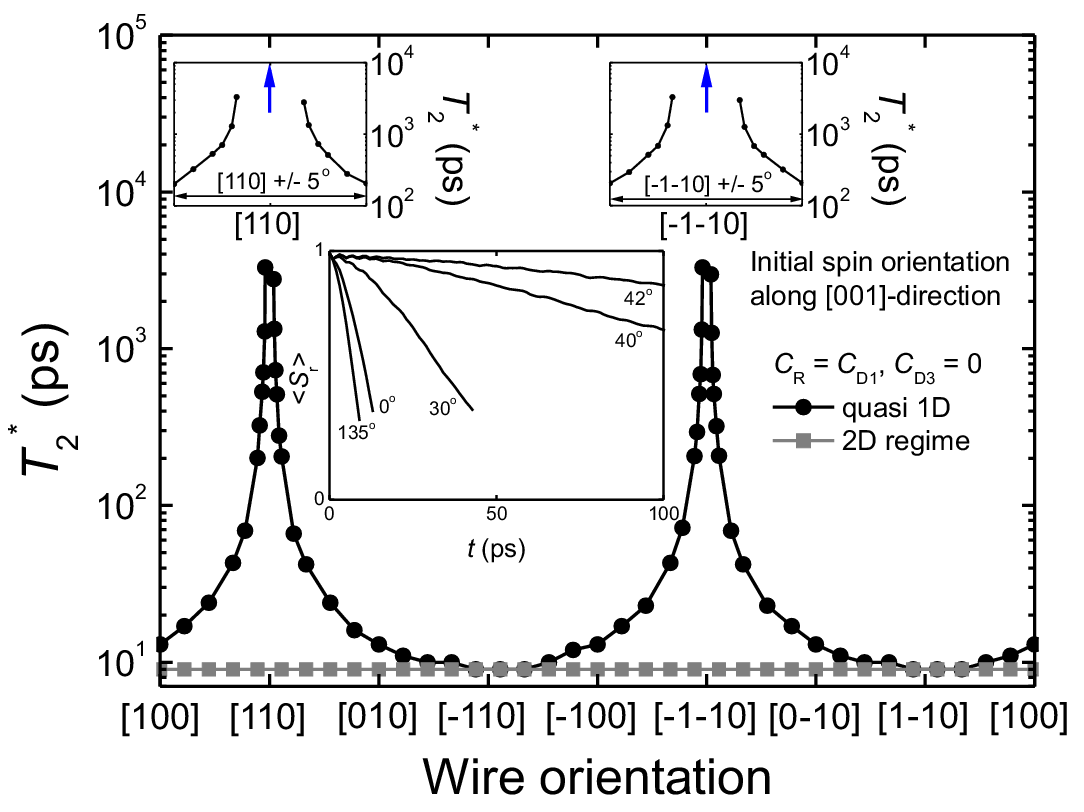}
\end{center}
\caption{The spin-dephasing time $T_2^*$ of a spin ensemble is
plotted as a function of the wire orientation with respect to the
[100] lattice direction. The ensemble is initially oriented along
the [001] direction. $T_2^*$ is strongly enhanced for a quasi-1D
wire oriented in the [110] direction (black) as compared to an
ensemble in a 2D system (gray). The arrows (in the top insets,
horizontal axes span $\pm 5^{\circ}$) for data at 45$^{\circ}$ and
225$^{\circ}$ indicate that here $T_2^*$ is larger than could be
calculated ($C_R$ = $C_{D1}$ = $-1.57\cdot$10$^{-8}$ Tm, $C_{D3}$ =
0). Inset: Ensemble spin expectation value $\langle S_r \rangle$ as
a function of time for different wire orientations.} \label{5}
\end{figure}
The inset of Figure \ref{5} shows the temporal evolution for five
different orientations of the quasi-1D wire with respect to the
[100] direction. A distinct anisotropy of the spin dephasing times
is observed. The peak value of $T_2^*$ is reached when the wire is
oriented exactly along the [110] direction. It is found to be in
excess of 10$^{4}$ ps, but the exact value could not be calculated
in a reasonable computation time. Yet, a deviation of only 5$^\circ$
in the wire orientation with respect to the [110] direction leads to
a reduction of $T_2^*$ of more than an order of magnitude towards a
value of about 200 ps. For angles close to [110] $\pm$ 15$^\circ$
the spin-dephasing time drops already to 43 ps and it reaches a
minimum of 9 ps for wires oriented along the [$\overline{1}$10] and
[1$\overline{1}$0] direction. A detailed summary of these findings
is presented in Figure~\ref{5} where the extracted spin-dephasing
times are plotted as a function of the wire orientation. This plot
reveals the strong anisotropy of $T_2^*$ with respect to the crystal
axes.
\\
The anisotropy of $T_2^*$ is directly connected to the
motion of single electrons within the ensemble. Specular
edge scattering implies that electrons with a solely
transverse momentum component to the wire orientation
(traveling less than the spin precession length between
scatter events) almost do not contribute to the
spin-dephasing because of motional narrowing. Only
electrons with a strong momentum component longitudinal to
the wire orientation are contributing to the dephasing of
the spin ensemble. This results in the strong enhancement
of $T_2^*$ for wires in the [110] direction.
\\
Spin-dephasing times for non-confined spin ensembles (wire
width taken much wider than the mean free path and spin
precession length) are also calculated (gray,
Figure~\ref{5}). In contrast to the quasi-1D wire case, no
spin-dephasing anisotropy is found. We investigated the
crossover from 2D to quasi 1D behavior for a wide range of
values for the spin precession length and mean free path
(with respect to the wire width), and found that the spin
precession length is the crucial length scale which is
governing this crossover.
\\
Next, we discuss whether this distinct spin-dephasing
anisotropy in quasi-1D wires (with a strong enhancement of
$T_2^*$ for wires in the [110] direction as a finger print)
can be maintained under more realistic circumstances. First
we study the influence of adding the cubic Dresselhaus term
on the enhancement of $T_2^*$. Secondly, the influence of
deviating strength of Rashba and linear Dresselhaus fields
will be discussed. To avoid the difficulty that $T_2^{*}$
cannot be calculated for 45$^\circ$ with respect to the
[100] direction within reasonable calculation time, we take
43$^\circ$ as a test case, as this already shows a very
strong $T_2^{*}$ enhancement, while $T_2^{*}$ is limited to
nanoseconds. Hence, the calculated spin-dephasing times in
the following part can be seen as lower bounds. $T_2^*$ for
the exact [110] direction is expected to be distinctively
higher.
\\
The influence of the additional cubic Dresselhaus term on
the spin dephasing time in the quasi-1D wire is presented
in Figure~\ref{6} (a). Again, the 2D case is plotted as a
reference (gray). Without the cubic Dresselhaus term the
calculation results in a $T_2^*$ of 1.3 ns. However, the
spin-dephasing time is decreasing rapidly with increasing
Dresselhaus parameter. For $C_{D3}$ = -7$\cdot$10$^{-25}$
Tm$^{3}$, $T_2^{*}$ is already reduced to 78 ps. To
estimate whether the spin-dephasing anisotropy is still
present in a more realistic situation, experimentally
deduced parameters are applied for comparison with our
calculations. In \cite{Miller2003} a value of $C_{D3}$ =
-1.18$\cdot$10$^{-24}$ Tm$^{3}$ is evaluated which results
in a $T_2^*$ of only 40 ps. A value which is less than an
order of magnitude higher than the value calculated for the
2D case. This points to the conclusion that the cubic
Dresselhaus term nearly annihilates the spin-dephasing
anisotropy. However, $C_{D3}$ depends strongly on the
electron density of the system. For samples with lower
densities $C_{D3}$ is orders of magnitude smaller
\cite{Miller2003, Jusserand1995, Jusserand1993}.
Considering those values, it turns out, that there is a
much weaker decay of the peak value of the spin-dephasing
time, even when the cubic Dresselhaus term is included.
\begin{figure}[htb]
\begin{center}
\includegraphics[width=8cm]{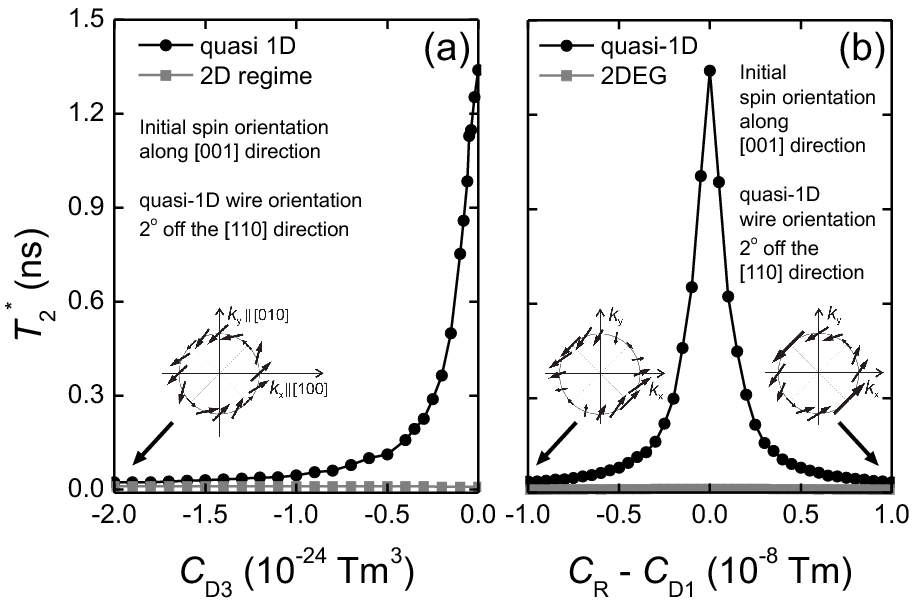}
\end{center}
\caption{Data shows how the $T_2^*$ enhancement of Figure~\ref{5}
reduces when the cubic Dresselhaus term is added and when the
symmetry between $C_R$ and $C_{D1}$ is lifted. The initial spin
state is chosen to be along [001] direction and the quasi-1D wire is
set at 43$^{\circ}$ with respect to the [100] direction. The case of
43$^{\circ}$ wire orientation (rather than 45$^{\circ}$) avoids the
need to deal in calculations with extremely long $T_2^*$ for the
symmetric case of $C_R$ = $C_{D1}$ and $C_{D3}$ $\approx$ 0, while
still clearly showing the 1D $T_2^*$ enhancement. (a) The effect of
the cubic Dresselhaus term on $T_2^*$ is plotted for the 2D case
(gray) and for the quasi-1D case (black) ($C_R$ = $C_{D1}$ =
-1.57$\cdot$10$^{-8}$ Tm). (b) $T_2^*$ is plotted here against the
difference $C_R$ - $C_{D1}$, at the fixed value of $C_{D1}$ =
-1.57$\cdot$10${^{-8}}$ Tm for the 2D case (gray) and for the
quasi-1D case (black) ($C_{D3}$ = 0).} \label{6}
\end{figure}
Finally, $T_2^*$ is investigated for deviating Rashba and linear
Dresselhaus contributions. This dependence is summarized in
Figure~\ref{6} (b) where $T_2^*$ is plotted as a function of the
difference in strength of the Rashba and linear Dresselhaus
parameter, $(C_{R} - C_{D1})$, in the interval
$\pm$1$\cdot$10$^{-8}$ Tm for $C_{D1}$ fixed at $-1.57
\cdot$10$^{-8}$ Tm. At $C_{R} - C_{D1} = 0$ this results in the
previously calculated $T_2^*$ of around 1.3 ns. With either
increasing $|C_{R}|$ or increasing $|C_{D1}|$, $T_2^*$ is decaying
equally fast. $C_R$ and $C_{D1}$ taken from \cite{Miller2003} result
in a difference $C_{R} - C_{D1}$ of 0.4$\cdot$10$^{-8}$ Tm. For this
value our calculated $T_2^*$ is already considerably reduced to
about 110 ps. However, this spin dephasing time is still an order of
magnitude higher than the one resulting from the 2D case and in
addition, $C_R$ can be tuned with a gate or heterostructure design
to equalize it to the linear Dresselhaus field. Therefore, in
summary, it can be stated that the spin-dephasing anisotropy which
is very distinct for $C_{R}$ and $C_{D1}$ exactly equal can still
prevail under less ideal conditions.

\section{\label{Conclusion}Conclusion}
A useful numerical tool is developed for studying spin-dephasing in
device structures due to the D'yakonov-Perel' spin-dephasing
mechanism. The Rashba, linear and cubic Dresselhaus contributions
can be taken into account. With this tool is was demonstrated that
quasi-1D wires (narrower than the spin precession length, but with
diffusive 2D motion for the electron ensemble) can show very clear
signatures of spin-dephasing anisotropy, with a strong suppression
of spin dephasing for wires in the [110] crystal direction.

\begin{acknowledgments}
We acknowledge financial support from the Dutch Foundation for
Fundamental Research on Matter (FOM), the Netherlands Organization
for Scientific Research (NWO) and one of us (T. Last) acknowledges
financial support by the Dutch Nanotechnology Program NanoNed.
\end{acknowledgments}

\end{document}